\documentclass{article}
\usepackage{ijcai21}

\usepackage{times}

\usepackage{soul}
\usepackage{url}
\usepackage[hidelinks]{hyperref}
\usepackage[utf8]{inputenc}
\usepackage[small]{caption}
\usepackage{graphicx}
\usepackage{amssymb}
\usepackage{amsmath}
\usepackage[thmmarks,amsmath]{ntheorem}
\usepackage{booktabs}
\usepackage[ruled,vlined]{algorithm2e}
\usepackage{diagbox}
\usepackage{subfigure}
\usepackage{tkz-euclide}
\usetikzlibrary{arrows,patterns,calc}
\usepackage{lipsum}
\usepackage[margin=1in]{geometry}

\newtheorem{theorem}{Theorem}
\newtheorem{lemma}{Lemma}
\newtheorem{corollary}{Corollary}
\newtheorem{definition}{Definition}

\newtheorem{proposition}{Proposition}
\newtheorem{proofLemma}{Proof of Lemma}
\newtheorem{proofCorollary}{Proof of Corollary}
\newtheorem{proofTheorem}{Proof of Theorem}
\newtheorem{proofProposition}{Proof of Proposition}
\newcommand*{\QEDB}{\null\nobreak\hfill\ensuremath{\square}}%
\SetKwInput{KwInput}{Input}                
\SetKwInput{KwOutput}{Output}              

\title{A Contract Theory based Incentive Mechanism for Federated Learning}

\author{
Mengmeng Tian$^1$\and
Yuxin Chen$^1$\and
Yuan Liu$^1$\footnote{Corresponding Author}\and
Zehui Xiong$^{2}$\and
Cyril~~Leung$^3$ \and
Chunyan~~Miao$^{34}$\\
\affiliations
$^1$Software College, Northeastern University, China\\
$^2$Pillar of Information Systems Technology and Design, Singapore University of Technology Design, Singapore.\\
$^3$Joint NTU-UBC Research Centre of Excellence
in Active Living for the Elderly (LILY), Nanyang Technological University, Singapore\\
$^4$School of Computer Science and Engineering, Nanyang Technological University, Singapore\\
\emails
\{1901270,2001236\}@stu.neu.edu.cn,liuyuan@swc.neu.edu.cn, zehui\_xiong@sutd.edu.sg, cleung@ece.ubc.ca,
ascymiao@ntu.edu.sg
}

\begin{document}

\maketitle

\begin{abstract}
  Federated learning (FL) serves as a data privacy-preserved machine learning paradigm, and realizes the collaborative model trained by distributed clients. To accomplish an FL task, the task publisher needs to pay financial incentives to the FL server and FL server offloads the task to the contributing FL clients. It is challenging to design proper incentives for the FL clients due to the fact that the task is privately trained by the clients. This paper aims to propose a contract theory based FL task training model towards minimizing incentive budget subject to clients being individually rational (IR) and incentive compatible (IC) in each FL training round. We design a two-dimensional contract model by formally defining two private types of clients, namely data quality and computation effort.  To effectively aggregate the trained models, a contract-based aggregator is proposed. We analyze the feasible and optimal contract solutions to the proposed contract model. 
  Experimental results show that the generalization accuracy of the FL tasks can be improved by the proposed incentive mechanism where contract-based aggregation is applied.
  
\end{abstract}

\section{Introduction}
With the ubiquitous adoption of Internet connected smart devices and applications,  the volumes of private data is growing in an unprecedented speed.  In a traditional data driven machine learning paradigm,  such large volumes of data are stored and analyzed on a third-party cloud server benefiting from its advantages of computing and storage capacities. However, with the data privacy issue ever-rising in both academics and industry, this centralized paradigm becomes unpractical. In this context, federated learning (FL) was proposed in \cite{DBLP:conf/aistats/McMahanMRHA17,DBLP:conf/ccs/BonawitzIKMMPRS17} and it has emerged as a potential solution in order to address this privacy issue, where the private data is stored and used to train a model at end-devices locally.  

In a classical FL framework, an FL server posts a target model referred to as an FL task to be collaboratively trained by distributed FL clients. To attract FL clients actively participating in training the FL task and compensate their efforts in executing the task, the FL server is necessary to offer sufficient economic incentives for FL clients \cite{DBLP:journals/corr/abs-1908-03092}. There are many studies that investigate the design of incentive mechanism for federal learning \cite{DBLP:conf/globecom/HuG20,DBLP:conf/ithings/FengN0KL19,DBLP:journals/cm/KhanPTSHNH20}, such as contract theory based mechanisms \cite{DBLP:journals/iotj/KangXNXZ19,contract2}.
In these existing contract based solutions for federated learning, the data quality towards improving model generalization accuracy~\cite{generalization} is rarely discussed, which is an essential performance metric of a deep learning model. What's more, all these contract models only studies adverse selection issue where the FL server offers tasks and contracts to be chosen by clients according to their types. The clients may not always put their efforts in executing FL tasks resulting in moral hazard issue. In this study, we aim to design a multi-dimensional contract considering clients' data quality in model generation performance and effort willingness.  The main contributions of this study are summarized as follows. 
\begin{figure*}[!ht]
	\centering
	\includegraphics[trim={30mm 182mm 30mm 10mm},clip,width=1\textwidth]{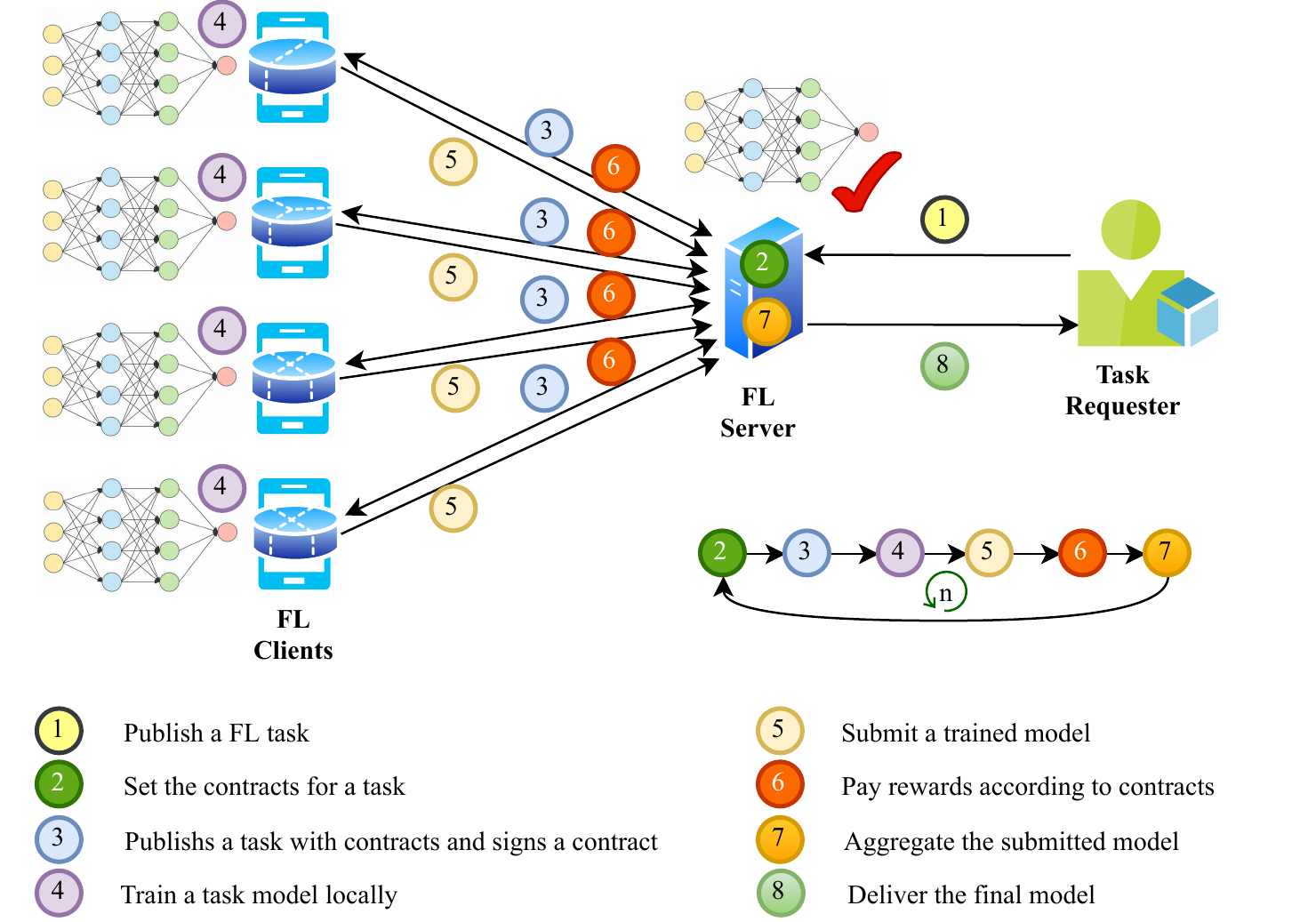}
	\caption{Overview of the proposed contract theory based federated learning procedure}
	\label{fig:model_train}
\end{figure*}
\begin{itemize}
     \item A contract theory based federated learning procedure is proposed, including 8 main processes, to support contract base incentive mechanisms for FL platforms.    
    \item A multi-dimensional contract model is designed by formally constructing the two private types of FL clients, i.e., generation type and effort willingness. The utilities of clients and FL server are formalized to solve an optimal contract solution. 
    \item A contract based aggregation scheme is designed to improve the model generalization accuracy. Experimental results based on MNIST and CIFAR-10 dataset show that the proposed contract based incentives and aggregation scheme outperforms other schemes in a single FL training  round.

\end{itemize}

\section{System Model}
We consider a classical federated learning platform where an FL task is proposed by a task requester and delegated to a trusted FL server, and the FL server coordinates the FL task distributedly trained by a set of FL clients. The FL clients participate in the FL task training under incomplete information where FL clients privately train the task model according to their private type and FL server cannot observe the clients' behaviors or private types but be aware of the private type distribution. To effectively incentivize  FL clients to execute tasks, we propose a contract theory based federated learning procedure, as shown in Figure \ref{fig:model_train}.  It includes eight main processes. Specifically, a task requester posts a model task to FL server,  and the server calculates a set of contracts for the task. The server then publishes the task among the client network and clients can choose to sign a contract through registering to the task according to the chosen contract. The clients then take efforts in training the task model based on their private datasets. Upon a qualified local model is trained, a client can submit the model to the FL server and the server pays clients according to the corresponding contracts. The server then aggregates the submitted model according to a contract based aggregation scheme. In the following, we first formulate the proposed contract, and then specify the utility functions of clients and server, and finally introduce the design of contract based aggregation. 

\subsection{Contract based Federated Learning}
In the proposed system, FL clients have private local data
and the FL server cannot predict clients' behavior. We aim to design a contract mechanism to elicit such private information.  

Suppose there are $I$ types of clients whose data coverage quality types are sorted in an ascending order:  $\theta_1\le \ldots \le \theta_i \le \ldots \le \theta_I$.  For each FL iteration round, the FL server needs to propose a  contract set ${\Phi}=\{\phi_i=(f_i,R_i(f_i))|i\in \{1,\ldots,I\}\}$  to specify the relationship between clients' rewards and registration fee for each client type, where $f_i$ is the registration fee for clients in $i-$th type to accept a task and $R_i(f_i)$ is the corresponding rewards. The design of $f_i$ aims to ensure the participation of rational clients before accepting a contact and the clients will not participate if they are unable to execute the task.  Then the server broadcasts the contract set among the clients, and each client signs a contract according to its type.  The clients then start training the model based on their local data and finally submit the trained model within the time requirement. The server makes a test about the generalization accuracy of submitted models. The clients are rewarded with $R_i(f_i)$ if the model passes the test by reaching the corresponding generalization accuracy $M_i$, and nothing otherwise. For the clients not rewarded, their registration fee will not be returned and used as the clients' penalty for breach of contract.  


It is worth noting that the model aggregation of the server process can be iterated for $n$ rounds, and the contract set should be set for each round. For a consecutively following round of a task, the generalization accuracy threshold $M_i$ should be set with a marginal increase. In this paper, without loss of generality, we design the contracts for a single aggregation round. 

Next, we formalize the types of clients and their utility. 

\subsection{Data Coverage Quality with Adverse Selection}	  
In the context of federated learning, to protect the data privacy of  FL clients, the clients are required to provide the trained models to the server instead of directly sharing their local data. Because of the asymmetric information,  the quality of model uploaded by clients can not be verified, which raises the adverse selection issue \cite{akerlof1970the}. We then model the client private type based on the local data quality in the aspect of model generalization capacity \cite{generalization}. 


Suppose the feature space with $d$ dimension is denoted by
	$D=\left[0,1\right]^{d}$ which is a unit space
and a subspace $\mathcal{A}\in D$.  Let $\mu(\mathcal{A})$ be the probability that a random sample in $D$ covered by $\mathcal{A}$ and $\mu(\mathcal{A})=1$ when $\mathcal{A}=D$. 

\begin{definition}[ $\epsilon$-Data Coverage] With a certain radius $\epsilon$, the coverage of a data set $\mathcal{A}$ consisting of samples $x_1,...x_m$ is measured by 
	\begin{equation}
	\mu(\mathcal{A},\epsilon)=D \cap \cup_{x_{i} \in \mathcal{A}} B(x_{i}, \epsilon) 
\end{equation}
where $B(x_{i}, \epsilon)$ is a open ball space centered at $x_{i}$ with radius $\epsilon$.
\end{definition} 
Suppose the data space is a Euclidean space, then the range of $\epsilon$ is $[0,\sqrt{d}]$. 
\begin{definition}[Data Coverage Quality] The data coverage quality of a local dataset $\mathcal{A}$ is denoted by $\theta(\mathcal{A})$ to be measured as the expected coverage expected coverage
\begin{equation}
	\theta(\mathcal{A})=\frac{1}{\sqrt{d}}\int_{0}^{\sqrt{d}} \mu(\mathcal{A},\epsilon) {\rm d}\epsilon 
\end{equation}
\end{definition} 



We consider a concrete set of data coverage quality denoted by $\Theta=\{\theta_1,...\theta_{I}\}$ with $I$ types, where the clients with $\theta(\mathcal{A})\in[\frac{i-1}{I},\frac{i}{I}]$ belong to type $i$.

\subsection{Training Willingness with Moral Hazard}
The client node consumes its local resource to complete an FL task and the efforts in training is a piece of private information, bearing moral hazard issue \cite{holmstrom1979moral}. We model the second type of a client as training willingness characterized by its efforts in training an FL task. 
\begin{definition}[Training Willingness] 
The training willingness of a client, denoted by $e\in [0,1]$, is the extent of the client taking their efforts in a task training. 
\end{definition}

The training cost of a client is especially determined by its training willingness, which is measured by convex function\cite{contract4,DBLP:conf/globecom/YuZL16} as
\begin{equation}\label{EffortOfNode}
	\Phi(e)=\frac{c}{2}e^2
\end{equation} where $c$ represents the unit cost in a given task training environment, such as IoT devices, smart mobile, PC, or server. Without loss of generality, we denote the training willingness of clients in $i$-th quality type as $e_i$.

\subsection{Utility of Client}
With a two-dimensional private information $\theta_i$ and $e_i$, the cost of a client in the $i$-th type is
\begin{equation}\label{CostOfNode}
	C(\theta_i,e_i) = f_{i} +\frac{c}{2}e_{i}^{2} 
\end{equation}

The client is rewarded with $R_i(f_i)$ if the trained model pass the generalization test benchmark $M_i$. The probability of passing the test is determined by the data coverage quality and training willingness \cite{YanruZhang}. Therefore, the utility of the client is 

\begin{equation}\label{UtilityOfNodeI}
U_{i}=\theta_{i}e_{i}R_{i} - f_{i} - \frac{c}{2}e_{i}^2.
\end{equation}

\subsection{Utility of Server}
The local model uploaded by the client of type $i$ will generate a revenue for the FL server, denoted by $G(M_{i})$ satisfying $G'(M) > 0$ and $G''(M) > 0$. Therefore, the utility of the server from enrolling client in type $i$ is 
\begin{equation}\label{UtilityOfPublisherI}
	U_{s}^i=f_{i} + \theta_{i}e_{i}(G(M_{i})-R_{i})
\end{equation} 

Given the type distribution of clients $\{\beta_{i}\}$ with $i\in\{1,...I\}$ and $\sum_{i=1}^{I}\beta_i=1$, the expected utility of the server is 
\begin{equation}\label{UtilityOfPublisher}
U_{s} =\sum_{i=1}^{I}\beta_{i}U_s^i= \sum_{i=1}^{I}\beta_{i}(f_{i} + \theta_{i}e_{i}(G(M_{i})-R_{i}))
\end{equation}

\subsection{Contract Optimization Problem}

The contract optimization problem is formalized as 
\begin{equation}\label{ProblemFormulation}
	\begin{split}
	\max &\sum_{i=1}^{I}\beta_{i}(f_{i} + \theta_{i}e_{i}(G(M_{i})-r_{i})) \\
	s.t. & \\
	(IR) &\theta_{i}e_{i}R_{i} - f_{i} - \frac{c}{2}e_{i}^2 \ge 0\\
	(IC) &\theta_{i}e_{i}R_{i} - f_{i} - \frac{c}{2}e_{i}^2 \ge \theta_{i}e_{i}^{j}R_{j} - f_{j} - \frac{c}{2}(e_{i}^{j})^ 2 \\
	&\forall j\neq i, \ i,j \in \{1, \dots, n\} 
	\end{split}
\end{equation}
where $e^{j}_{i}$ denotes the effort of type $\theta_{i}$ when selecting contract ($f_j,R_j$). 

The first constraint ensures that each client can achieve non-negative utility, which is also regarded as individual rationality property (IR). The second constraint aims to ensure that each client can achieve their maximal utility by choosing the contract corresponding to their truthful type, which is regarded as incentive compatibility property (IC).


\subsection{Contract-Based Model Aggregation}


 With a set of submitted models, the server should aggregate the models based on their chosen contracts for the sake of better model generalization performance. Suppose the total rewards paid by a server in a round is $R_{total}$, then the weight assigned for a model trained by a client in type $i$ is calculated according to Eq.\eqref{eq:contractWeight}.
\begin{equation}\label{eq:contractWeight}
	w_{i} = \frac{R_i}{R_{total}}
\end{equation}

\section{Optimal Contract Solution}
In this section, we solve the optimal contract solution to the problem defined in Eq.\eqref{ProblemFormulation}. We first solve the optimal effort willingness made by clients and then calculate the contract solution by maximizing the server utility. 

Given the utility of a client in $i$-th type in Eq.\eqref{UtilityOfNodeI}, we compute the first order derivative with respect to its effort willingness and we obtain 
\begin{equation}
	\frac{\partial U_i}{\partial e_i}=\theta_i R_i-c e_i
\end{equation}
A rational client node should always maximize its utility by making the optimal willingness which is denoted by $\hat{e}_i$ and $\hat{e}_{i}^{j}$ in choosing contract $\phi_i=(f_i,R_i(f_i))$ and contract $\phi_j=(f_j,R_j(f_j))$. Thus, 

\begin{equation}\label{OptimalEffort}
	\hat{e}_{i}=\frac{1}{c}\theta_{i}R_{i}\quad, \quad \hat{e}_{i}^{j}=\frac{1}{c}\theta_{i}R_{j}
\end{equation}

According to Eq.\eqref{OptimalEffort}, we can know that a client's  willingness is positively determined by the data quality and the chosen contract reward. 

Bring Eq.\eqref{OptimalEffort} into Eq.\eqref{ProblemFormulation}, the objective function is updated as follows.
\begin{equation}\label{PrblemWithoutE}
	\begin{split}
		\max &\sum_{i=1}^{I}\beta_{i}(f_{i} + \frac{1}{c}\theta_{i}^{2}R_{i}(G(M_{i})-R_{i})) \\
		s.t. &\\
		(IR) \ & \frac{1}{2c}(\theta_{i}R_{i})^{2} - f_{i} \ge 0,\\
		(IC) \ & \frac{1}{2c}(\theta_{i}F_{i})^{2} - f_{i} \ge \frac{1}{2c}(\theta_{i}R_{j})^{2} - f_{j}, \\
		&\forall j\neq i, \ i,j \in \{1, \dots, n\} 
	\end{split}
\end{equation} 

Next, we will solve the optimal contract solution for each type $\theta_i$. Some important conditions will be derived.

\begin{lemma}[Monotonicity between $\theta$ and $R$]
	For any feasible contract ($f_{i}, R_{i}$), $R_{i} \ge R_{j} \Leftrightarrow \theta_{i} \ge \theta_{j}$.
\end{lemma}

\begin{proofLemma}
For clients of type $\theta_i$ and $\theta_j$, the following two IC constrains should be satisfied
\begin{gather}
\frac{1}{2c}(\theta_{i}R_{i})^{2} - f_{i} \ge \frac{1}{2c}(\theta_{i}R_{j})^{2} - f_{j} \label{eq:b1}\\
\frac{1}{2c}(\theta_{j}R_{j})^{2} - f_{j} \ge \frac{1}{2c}(\theta_{j}R_{i})^{2} - f_{i} \label{eq:b2}
\end{gather}By adding the above two inequalities, we have
\begin{align}
&(\theta_{i}^{2}-\theta_{j}^{2})(R_{i}^{2}-R_{j}^{2})\ge 0 \\
\Rightarrow &(\theta_{i}-\theta_{j})(R_{i}-R_{j}) \ge 0
\end{align} for any $\theta_{i},\theta_{j} > 0$ and $R_{i}, R_{j} > 0$. \QEDB
\end{proofLemma} 

Lemma 1 imply that a client with a higher type $\theta$ fit for a higher rewards $R$. Thus, the contract rewards should follow the order $R_{1} < \dots< R_{I}$ with $\theta_{1} < \dots  < \theta_{I}$.

\begin{lemma}[Monotonicity between $R$ and $f$]
	For any feasible contract ($f_{i}, R_{i}$), $R_{i} \ge R_{j} \Leftrightarrow f_{i} \ge f_{j}$. 
\end{lemma}

\begin{proofLemma}
The IC constrain holds when a client in type $\theta_{i}$ chooses contract ($f_{i}, R_{i}$) compared with ($f_{j}, R_{j}$). 
\begin{align}
\frac{1}{2c}(\theta_{i}R_{i})^{2} - f_{i} \ge \frac{1}{2c}(\theta_{i}R_{j})^{2} - f_{j} \notag \\
\Rightarrow \left\{ \begin{array}{rl} f_{i} - f_{j}& \leq \frac{\theta_{i}^{2}}{2c}(R_{i}^{2}-R_{j}^{2}) \\
 f_{j} - f_{i} &\ge \frac{\theta_{j}^{2}}{2c}(R_{j}^{2}-R_{i}^{2}) \label{eq:b4}\end{array} \right.\end{align}
According to Eq.\eqref{eq:b4}, if  $f_{i} \ge f_{j}$, then we have $R_{i} \ge R_{j}$, and vice verse. \QEDB
\end{proofLemma}

Lemma 2 shows that $R$ and $f$ have the same trend, namely $R_{1} < \dots < R_{I}$ with $f_{1} < \dots < f_{I}$.

\begin{corollary}[Monotonicity between $f$ and $\theta$]
	For any feasible contract ($f_{i}, R_{i}$), $f_{i} \ge f_{j} \Leftrightarrow \theta_{i} \ge \theta_{j}$. 
\end{corollary}
\begin{proofCorollary}
According to Lemma 1 and Lemma 2, both $\theta$ and $f$ monotonically increase with $R$. Thus, we can derive that the positive correlation between $\theta$ and $f$.  \QEDB
\end{proofCorollary}

The above lemmas and corollary shows the monotonicity properties. 
Next, we will find the optimal contract by reducing the IR constrains and IC constrains. 

\begin{theorem}[IR transitivity]
	All the IR constrains can be satisfied if the constrain of  $\theta_{1}$ is satisfied.   
\end{theorem}
\begin{proofTheorem}
For any client in type $i \in \{1, \dots, I\} $ and $i \ge 1$, we have
\begin{equation}
\begin{split}
U_{i}=\frac{1}{2c}(\theta_{i}R_{i})^{2} - f_{i} \ge \frac{1}{2c}(\theta_{i}R_{1})^{2} - f_{1} \\
\ge \frac{1}{2c}(\theta_{1}R_{1})^{2} - f_{1}=U_{1} 
\end{split}
\end{equation} and its utility is monotonous. \QEDB
\end{proofTheorem}

\begin{theorem}[Tight IC Constrain]
	The following IC constrain is sufficient for client in type $\theta_{i}$ to achieve its maximal utility.  
	\begin{equation}\label{eq:IC}
		\frac{1}{2c}(\theta_{i}R_{i})^{2} - f_{i} = \frac{1}{2c}(\theta_{i}R_{i-1})^{2} - f_{i-1}
	\end{equation} where $i\in \{2,...,I\}$.
\end{theorem}
\begin{proofTheorem}
The following proof is organized as three parts. First of all, we reduce the redundant IC constrains in two direction: client in type $\theta_{i}$ select the contract $\phi_{i+1}$ and $\phi_{i-1}$ respectively. And then, All redundant constraints will be eliminated, leaving only tight constraints \eqref{eq:IC}.
\end{proofTheorem}

\noindent1) Downward Selection:
\begin{gather}
\frac{1}{2c}(\theta_{i + 1}R_{i + 1})^{2} - f_{i + 1} \ge \frac{1}{2c}(\theta_{i + 1}R_{i})^{2} - f_{i} \label{eq:a1}\\
\frac{1}{2c}(\theta_{i}R_{i})^{2} - f_{i} \ge \frac{1}{2c}(\theta_{i}R_{i-1})^{2} - f_{i-1} \label{eq:a2}
\end{gather}
Transfer \eqref{eq:a1} and \eqref{eq:a2},we have
\begin{align}
f_{i+1} - f_{i} &\le \frac{1}{2c}\theta_{i + 1}^{2}(R_{i + 1}^{2} - R_{i}^{2}) \label{eq:a3}\\
f_{i} - f_{i-1} &\le \frac{1}{2c}\theta_{i}^{2}(R_{i}^{2} - R_{i-1}^{2}) \label{eq:a41}
\end{align} Since $\theta_{i} < \theta_{i + 1}$,  according to \eqref{eq:a41} we have 
\begin{align}
	\frac{1}{2c}\theta_{i}^{2}(R_{i}^{2} - R_{i-1}^{2}) &\le \frac{1}{2c}\theta_{i+1}^{2}(R_{i}^{2} - R_{i-1}^{2}) \notag\\
	\Rightarrow f_{i} - f_{i-1}	&\le \frac{1}{2c}\theta_{i+1}^{2}(R_{i}^{2} - R_{i-1}^{2}) \label{eq:a4}
\end{align}
Add \eqref{eq:a3} and \eqref{eq:a4}, then we have 
\begin{gather}
f_{i+1} - f_{i - 1} \le \frac{1}{2c}\theta_{i+1}^{2}(R_{i+1}^{2} - R_{i-1}^{2}) \label{a5}\\
\Rightarrow \frac{1}{2c}(\theta_{i + 1}R_{i + 1})^{2} - f_{i + 1}
 \ge \frac{1}{2c}(\theta_{i + 1}R_{i-1})^{2} - f_{i-1} \notag
\end{gather}
Then, we can obtain all the downward IC constrains:
\begin{align*}
\frac{(\theta_{i + 1}R_{i + 1})^{2}}{2c}\! -\! f_{i + 1} \!
&\ge\! \frac{(\theta_{i+1}r_{i-1})^{2}}{2c}\! -\! f_{i-1}\! \ge \! \cdots \\
\ge 
\frac{(\theta_{i+1}r_{1})^{2}}{2c}\! -\! f_{1}
\end{align*}

Therefore, all the adjacent type of downward IC selection is enough to drive all the other downward selection. 

\noindent2) Upward Selection: Similarly, we can drive all the upward selections by adjacent selection like
%
\begin{align*}
\frac{1}{2c}(\theta_{i}R_{i})^{2}\! - \!f_{i}\! \ge\! \frac{1}{2c}(\theta_{i}R_{i+1})^{2}\!-\!f_{i+1}\! \ge\! \dots \\
\! \ge\! \frac{1}{2c}(\theta_{i}R_{I})^{2} \!-\! f_{I}
\end{align*}
Therefore, we can drive other IC constrains from adjacent upward type selection.

\noindent3) Tight IC constrain: From the above two steps, the IC constrain of adjacent is left. Then, we will further remove the redundant restrictions. The tight IC constrain can be replaced by adjacent downward IC constrain and monotonicity show as follows. 
\begin{proposition}
	The downward IC constrains can grantee the upward IC constrains.
\end{proposition}
\begin{proofProposition}
Accordind to the above steps 1) and 2), we have
\begin{align}
	\frac{1}{2c}(\theta_{i}R_{i})^{2} - f_{i} \ge \frac{1}{2c}(\theta_{i}R_{i-1})^{2} - f_{i-1} \label{eq:c1}\\
	\frac{1}{2c}(\theta_{i}R_{i})^{2} - f_{i} \ge \frac{1}{2c}(\theta_{i}R_{i+1})^{2} - f_{i+1} \label{eq:c2}
\end{align}
According to Eq.\eqref{eq:c2}, we have 
\begin{align}
	f_{i+1} \ge \frac{1}{2c}\theta_{i}^{2}(R_{i+1}^{2} - R_{i}^{2}) + f_{i}\\
    \Rightarrow f_{i} \ge \frac{1}{2c}\theta_{i-1}^{2}(R_{i}^{2} - R_{i-1}^{2}) + f_{i-1} \label{eq:c4}
\end{align} and according to Eq.\eqref{eq:c1}, we have
\begin{equation}
	f_{i} \le \frac{1}{2c}\theta_{i}^{2}(R_{i}^{2} - R_{i-1}^{2}) + f_{i-1} \label{eq:c5}
\end{equation} 
Comparing Eq.\eqref{eq:c4} and Eq.\eqref{eq:c5}, given $\theta_{i} > \theta_{i-1}$, we can derive that the server will achieve its maximal utility when $f_i$ is assigned with its maximal value. So the tight IC constrain is Eq. \eqref{eq:IC}. \QEDB
\end{proofProposition}  

Taking the tight constrains into the objective function in Eq. \eqref{PrblemWithoutE}, we can formulate the Lagrange function  to solve the final optimal contract:   
\begin{equation}
\begin{split}
\mathcal{L} \!=\!& \!\sum_{i=1}^{I}\left\{\left[\beta_{i}(f_{i} +  \frac{1}{c}\theta_{i}^{2}r_{i}(G(M_{i})-R_{i}))\right] + \right.\\ &\left.\lambda_i\left[\frac{1}{2c}(\theta_{i}R_{i})^{2}  \!- \!\frac{1}{2c}(\theta_{i}R_{i-1})^{2} \!-\! f_{i}\! +\! f_{i-1}\right]\right\}\\ +&\mu\left[\frac{1}{2c}(\theta_{1}R_{1})^{2} - f_{1}\right]
\end{split}
\end{equation}where $\lambda_i$ is Lagrange multiplier of IC constrain for $\theta_i$, and $\mu$ is Lagrange multiplier of IR constrain for $\theta_1$.

%

Thus the optimal value of $R$ is
\begin{equation}\notag
	R_{i}=G(M_{i}), \forall i \in \{1, \dots, I\}
\end{equation} and the optimal value of $f$ can be driven according the IR transitivity and IC transitivity.
\begin{align*}
f_{1} &= \frac{1}{2c}(\theta_{1}R_{1})^{2}\\
f_{i} &= \frac{1}{2c}(\theta_{i}R_{i})^{2} - \frac{1}{2c}(\theta_{i}R_{i-1})^{2} + f_{i-1},\forall i \in \{2, \dots, I\}
\end{align*}

The above solution is the optimal contract solution after relaxing the constrains. We need to further verify whether the solution satisfies monotonicity condition of $R$. If $R$ is not monotonic, the adjustment algorithm \cite{SpectrumTrading} can be applied. 

\begin{table*}[ht]
	\centering
	\caption{Contract and Client Settings Of MNIST Task}
	\label{tab:contractMNIST}
	\begin{tabular}{|l|l|l|l|l|l|l|l|l|l|l|}
		\hline
		\diagbox{Parameters}{Type Index} & $1$ & $2$ & $3$ & $4$ & $5$ & $6$ & $7$ & $8$ & $9$ & $10$ \\
		\hline
		Client Type $\theta_i$     & 0.790    &      0.795        &    0.800          &      0.805         &     0.810        &      0.815        &       0.820       &     0.825         &       0.830       &    0.835           \\
		\hline
		Client Data Size    & 1000    &      1500        &    2000          &  2500      &    3500         &     5000        &       6500      &     8500        &       12000       &     16000          \\
		\hline
		Optimal effort $e_i$  &       0.279     &     0.331         &      0.389        &      0.451        &         0.519     &      0.592       &        0.670      &    0.753          &      0.842        &    0.936           \\
		\hline
		Test Generalization $M_i$ &     0.230         &      0.250       &     0.270         &      0.290        &     0.310         &      0.330        &      0.350        &     0.370         &     0.390        &     0.410          \\
		\hline
	\end{tabular}
\end{table*}

\section{Experiment Results and Analysis}
In this section, we evaluate the proposed contract based incentive mechanism for FL in two classical datasets, i.e.  MNIST and CIFAR-10. The proposed model is compared with other two schemes to demonstrate the effectiveness of our model in the aspect of generalization accuracy.


\subsubsection{Experimental Settings}
There are 10 types of contract and 10 types of clients are set up to correspond to the corresponding contracts. The types of clients follow a uniform distribution, i.e., $\lambda_i = 0.1$. The contract and client settings are shown in Table \ref{tab:contractMNIST}. 

Based on the above settings, we complete the steps from 1 to 6 in Figure \ref{fig:model_train}. After the client uploading their local model, the server will test whether the model meets the corresponding test generalization benchmark $M$ for each client according to the chosen contract. Due to the fact that the data are heterogeneous and follow non-IID, the data of the server used for testing is a part of the whole dataset and chosen randomly in the whole data space, serving as a sampling test for prevention of fraud. 
At the same time, through experiments, we found that although our model training accuracy in the local environment is 93\% , it is only 46\% in the server test. This fully shows the test error caused by different data coverage between server and client. Therefore, if the test baseline $M$ is set too high, the test results of the model will deviate greatly. In this experiment, the $M$ we set is relatively low, as shown in Table \ref{tab:contractMNIST}. We require that with the increase of client types, our benchmark requirements increase by 2\%. 

The following three schemes are compared by setting different rewarding methods and aggregation protocols. 
\begin{itemize}
	\item Scheme-1: The clients are rewarded according to the proposed contract based solution and the server aggregates the submitted model according to the contracted based aggregation protocol. 
	\item Scheme-2: The clients are rewarded according to the proposed contract based solution and the server aggregates the submitted model according to FedAvg protocol~\cite{openProblem} in which the model aggregation weight is same for all clients. 
	\item Scheme-3: The clients are rewarded equally and the server aggregates the submitted model according to FedAvg protocol. 
\end{itemize}
Notice that we set the same rewarding scheme with average effort and reward of contract-based rewarding scheme for each client for fairness. 
In addition, we also adjust parameter $c$ to show the impact of service environment on model aggregation performance.

\subsubsection{Experimental Results}
\begin{figure}
	\centering  
	\includegraphics[width=0.45\textwidth]{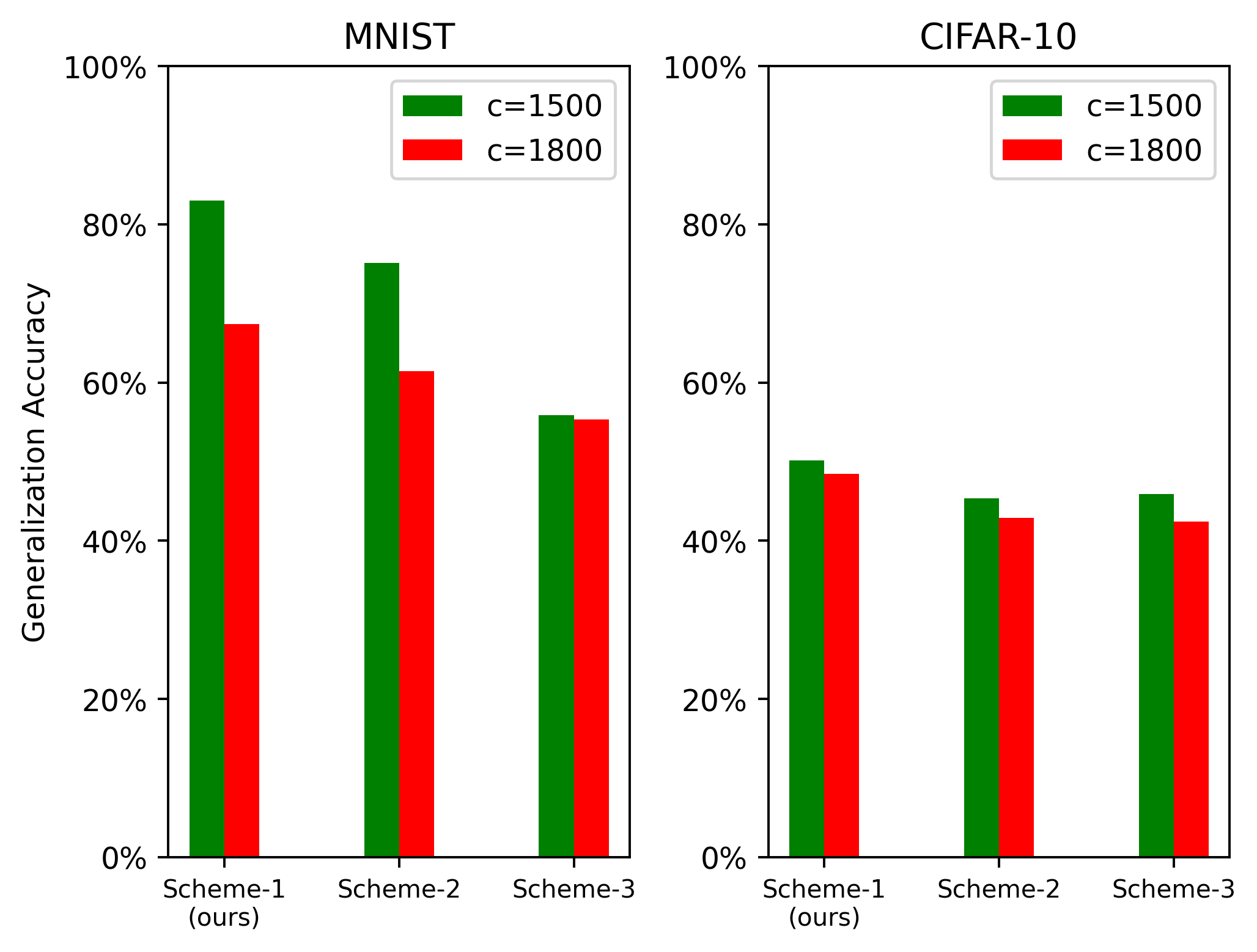}
	\caption{Generalization accuracy of the proposed scheme under different values of $c$} 	\label{fig:Gaccuarcy}  
\end{figure}

This generalization accuracy of the three schemes under two different parameter $c$ settings in two datasets is presented in Figure \ref{fig:Gaccuarcy}. In Figure \ref{fig:Gaccuarcy}, under the same parameter $c$, the proposed method, i.e. Scheme-1 shows the highest model generalization accuracy which is better that of Scheme-2 or Scheme-3. By using the contract-based incentive mechanism in Scheme-1 and Scheme-2, the clients work more hard and consistently perform more better than the fixed incentive in Scheme-1. The reason of better performance of Scheme-1over Scheme-2 is that Scheme-1 uses contract based aggregation which can set a higher weight for the model trained on a high quality data source. In addition,comparing the model accuracy under different parameter $c$, we can observe that the smaller parameter $c$ setting brings the better generalization accuracy, indicating that the clients with lower training cost is more likely to be incentivied to improve the model generalization performance. 

\section{Related Work}
At present, there are many researches on the  design of incentive mechanisms for federated learning based on game theoretical approaches \cite{DBLP:journals/comsur/WahabMOT21}.  
The existing incentive mechanisms can be mainly divided into two main categories: Stackelberg game-based and contract theory-based methods.  In the first categories, FL server offer a task associated with a price and clients choose a task and take efforts in training the task to achieve the pricing rewards \cite{DBLP:conf/globecom/HuG20,DBLP:conf/ithings/FengN0KL19,DBLP:journals/cm/KhanPTSHNH20}.   In \cite{DBLP:conf/globecom/HuG20}, a two-stage Stackelberg game is formalized for FL with private data, a Nash equilibrium (NE) is solved with the optimal privacy budget of clients and the optimal pricing scheme of server.  In \cite{DBLP:conf/ithings/FengN0KL19}, the interactions between model requester and mobile user are formalized as a Stackelberg game to analyze the NE composed by the optimal training data price and data size. In \cite{DBLP:journals/cm/KhanPTSHNH20}, a Stackelberg game based incentive mechanism is designed for FL clients in strategically set the local iterations and FL server maximizing task global accuracy. The shortcoming of the Stackelberg game based methods is that it can only consider a single dimension private strategical types. 

In the second category,   FL servers offer a set of contract items according to clients types where the contract models can be single-dimensional or multi-dimensional depending on the dimensions of the considered type. In \cite{contract2},  the client training quality type is formalized based and computation resources, and an optimal contract solution offers more rewards to clients with higher type values. Considering clients' communication delay and local training cost, \cite{DBLP:journals/jsac/DingFH21} introduces a contract-based incentive mechanism to maximize server aggregation accuracy and total payments, where the local training cost is related to the clients' network environment. \cite{contract4} considers both data quality and model computation resources. A most recent study \cite{DBLP:journals/jsac/DingFH21} investigates a two-dimensional contract model considering data quality in terms of data sizes and communication time types and analyze the optimal contract solutions in three scenarios: complete information, weakly incomplete information, and strongly incomplete information.  There are two limitations in the existing contract based methods:1) they neglect to consider the data quality in the aspects of improving model generalization accuracy, which will be studied as an important private type of clients; 2) they assume that the clients take their efforts in executing FL tasks, which bears the designed contracts with the moral hazard issue where their willingness may hinder the achievement of their optimal solution. 
\vspace{-2mm}
\section{Conclusion}
In this paper, we propose a contract based incentive mechanism for federated learning. A two-dimentional contract model is formally designed where we consider the client's data coverage quality and effort willingness. We also propose a contract based FL aggregation protocol. The optimal contract solution is theoretically analyzed. Finally the proposed incentive mechanism is experimentally evaluated and the results show that our contract based scheme achieves higher aggregation accuracy compared with the other two schemes. 
\vspace{-2mm}
\section*{Acknowledgements}
This work was supported by Alibaba Group through Alibaba Innovative Research (AIR) Program and Alibaba-NTU Singapore Joint Research Institute (JRI), Nanyang Technological University, Singapore; Key-Area Research and Development Program of Guangdong Province NO.2020B0101090005; National Natural Science Foundation of China under Grant No.62032013, and No.U20B2046; 111 Project (B16009); and the Fundamental Research Funds for the Central Universities N182410001.

\begin{scriptsize}
\bibliographystyle{named}
\bibliography{FLcontractShort.bib}	
\end{scriptsize}

\end{document}